# Modelling indirect interactions during failure spreading in a project activity network

Christos Ellinas[1,*]

[1] *Engineering Mathematics, University of Bristol, Bristol, UK*

[*] Corresponding author

E-mail: ce12183@bristol.ac.uk

**Abstract**

Spreading broadly refers to the notion of an entity propagating throughout a networked system via its interacting components. Evidence of its ubiquity and severity can be seen in a range of phenomena, from disease epidemics to financial systemic risk. In order to understand the dynamics of these critical phenomena, computational models map the probability of propagation as a function of direct exposure, typically in the form of pairwise interactions between components. By doing so, the important role of indirect interactions remains unexplored. In response, we develop a simple model that accounts for the effect of both direct and subsequent exposure, which we deploy in the novel context of failure propagation within a real-world engineering project. We show that subsequent exposure has a significant effect in key aspects, including the: (a) final spreading event size, (b) propagation rate, and (c) spreading event structure. In addition, we demonstrate the existence of 'hidden influentials' in large-scale spreading events, and evaluate the role of direct and subsequent exposure in their emergence. Given the evidence of the importance of subsequent exposure, our findings offer new insight on particular aspects that need to be included when modelling network dynamics in general, and spreading processes specifically.

**Introduction**

Recent years have witnessed a flurry of work on spreading processes[1,2], ranging from empirical expositions on the impact of such spreading (e.g. existence of large-scale spreading events[3-5], properties of 'super-spreaders'[6-10]) to methodological developments that map the underlying dynamics (e.g. spreading mechanisms[11-14], modelling frameworks[15-19]). Pairwise interactions are the perceived centrepiece in understanding the evolution of spreading processes[20], since they capture the *direct exposure* of each node to the prospect of switching its state, for example from 'non-affected' to 'affected'. Importantly, the impact of direct exposures can be supplemented by *subsequent* exposures which may arise due to global or local network features. Past work has focused on the impact of these indirect effects by considering the *global* topology of the network (e.g. distribution of shortest paths[21,22]) and how it is influenced by particular mechanisms (e.g. flow redistribution[23,24]). However, such indirect effects can also arise from *local*, non-trivial structures (e.g. particular network motifs such as the feed-forward loop[25,26]). Despite the intuitive importance of these local structures[26], little attention has been paid in evaluating their impact to the overall spreading process, largely due to the particularities of the spreading models typically deployed to study these processes.

In particular, spreading models can be classified to two broad categories[12,27], depending on the incorporated mechanisms: (i) epidemiological models, where the spreading process is viewed as an *independent* event across different pairs of nodes (e.g. the probability of node $i$ to affect its neighbour $j$ is independent of other interactions), and (ii) sociological models, where the spreading process is viewed as an *interdependent* event (e.g. the probability of node $i$ to affect its neighbour $j$ depends on the state of node's $j$ neighbours). Despite this distinction, both model categories are grounded on the same fundamental premise, in which direct exposure is the principal factor that determines the dynamics of the spread. Yet, the implication of this premise can be non-trivial, as subsequent exposures can interfere with the spreading process and affect key outcomes, even in simple examples like the one discussed below.

Consider the toy network in Fig. 1, where each node can switch states irreversibly from 'non-affected' to 'affected' with a given probability $p$. In the case of an epidemiological spreading model, this premise corresponds to the widely used 'Susceptible'-'Infected' model[28], where nodes and links correspond to individuals and infection pathways (e.g. social interactions), respectively. In this case, the ability of node $i$ to affect its immediate neighbours is assessed independently across all possible pairs (pair $k \to i$; pair $k \to l$; pair $k \to j$). At this point, node $j$ is directly exposed to the infection of node $k$, through the directed link from node $k$ to node $j$, and indirectly, through paths $k \to i \to j$ and $k \to l \to j$ (see Methods). As a result, the possibility for node $j$ to switch state is evaluated at three distinct points during the evolution of the spread (one direct link; two indirect paths), compared to the single evaluation that takes place in the case of node $i$ and node $l$ (due to their direct link). Hence, node $j$ is three times more likely to switch state, compared to node $i$ and $l$, despite that fact that the probability of them changing state is uniformly set. We can trace this effect to the implicit assumption that direct and subsequent exposures are of equal importance, which is appropriate when dealing with a typical disease spreading.

Fleshing out this assumption, the probability of node $j$ to switch states, either due to its contact with node $k$, or due to its contact with node $k$ (as part of path $k \to i \to j$), is exactly the same. This suggests that the underlying pathogen that drives the spread has remained unchanged (and therefore, the probability for node $k$ to infect its neighbours, including node $j$, is exactly the same as the probability of node $i$ to infect its own neighbours, including node $j$). However, consider an alternative context where the pathogen is replaced by a defect, which spreads across a network of activities, where nodes correspond to technical activities[29] (e.g. specify the design of an engineering component) and links to functional dependencies (e.g. activity $j$ can start if, and only if, its predecessor task $i$ finishes; i.e. manufacturing an engineering component can start only if its specifications have first been specified) [30,31]. In this case, the interpretation of direct and subsequent exposures is distinct, where activity $j$ may be particularly susceptible (or immune) to the particular defect that caused activity $i$ to fail. For example, consider the case where activity $k$ corresponds to 'specify component', activity $i$ to 'manufacture component' and activity $j$ to 'install component'. In the case where activity $k$ fails, the probability of affecting activity $j$ may be low (e.g. due to the availability of prefabricated, standardised components). Yet, if activity $i$ is affected by the failure of activity $k$, the probability of activity $j$ to be affected is now higher – in this case the defect propagating across the activity network has evolved from 'failure to specify' to 'failure to specify and manufacture'. Therefore, the fact that activity $j$ is additionally exposed to the defect in an indirect way (through node $i$ and node $l$; Fig. 1) directly interferes with the likelihood of activity $j$ changing its state; yet typical spreading models – like the aforementioned SI model – would assume that such subsequent exposure does not interfere.

In conjunction with the fact that activity networks have a high concentration of local, non-trivial structures[32] – and hence, experience pronounced levels of subsequent exposure (see Supplementary Fig. 1) - the likelihood of obtaining misleading results through the use of typical spreading models is high. Consequently, a range of decisions that may be driven by those results can be significantly affected, from estimating an appropriately sized contingency budget to developing appropriate (project) risk management framework(s)[31].

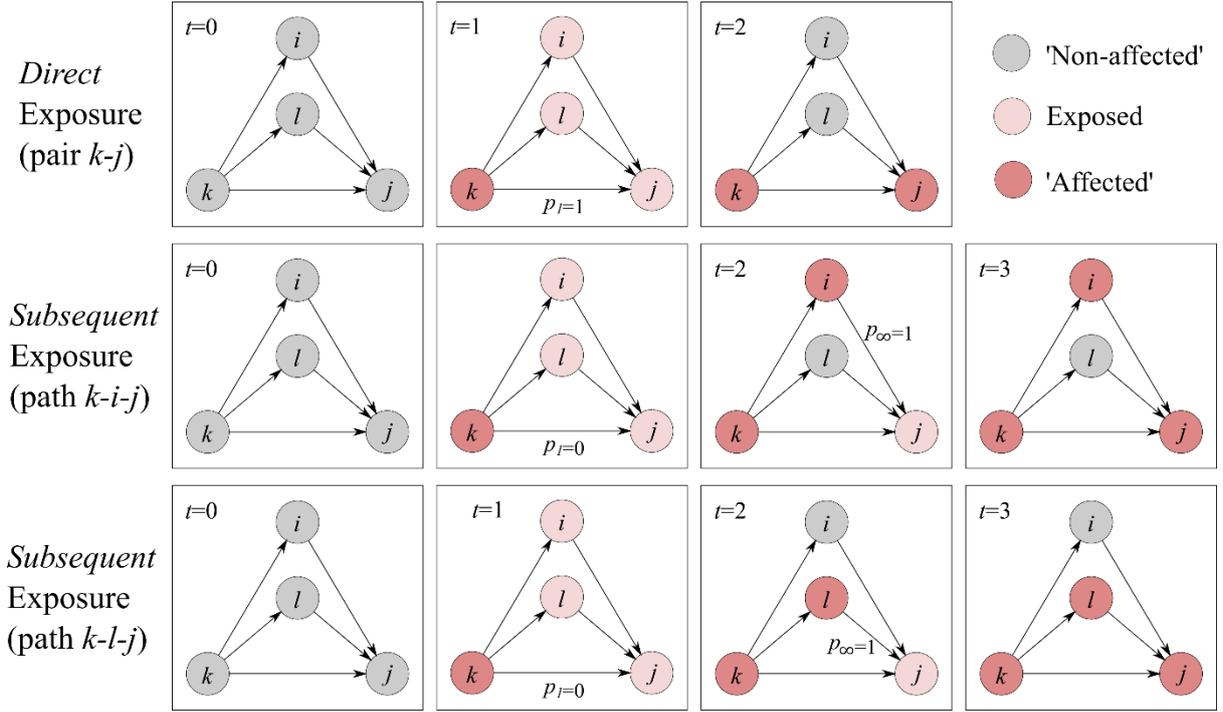

**Fig. 1:** Example to highlight the distinction between the direct and subsequent exposure of node $j$ to the failure of node $k$ (time runs from left to right). Top panel focuses on the direct case, where node $j$ is directly exposed to node $k$'s failure; rest illustrate the case where node $j$ is subsequently exposed, via node $i$ (middle panel) and node $l$ (bottom panel).

In response, we propose a simple spreading model that allows us to map the impact of direct and subsequent exposures (see Methods). It does so by disentangling the probability of a defect to spread ($p$), to two distinct components: $p_1$, which controls the probability of node $j$ being affected by the failure of its predecessor, node $k$ (Fig. 1, top panel) and $p_\infty$, which controls the probability of node $j$ being affected by additional, subsequent exposures to the failure of node $k$, through node $i$ (Fig. 1, middle panel) or node $l$ (Fig. 1, bottom panel). We distinguish between the two by keeping track of node $k$'s successor at time $t$, say node $j$, and assess whether node $j$ has been encountered before. If so, then node $j$ has been directly exposed to the failure of a *different* predecessor before $t$ (Fig. 1 middle/bottom panel at $t = 2$), and hence the probability of node $j$ failing is controlled by $p_\infty$; if node j is encountered for the first time, the probability of is controlled by $p_1$ (Fig. 1 middle/bottom panel at $t = 1$).

We formulate two models, $M_0$ where $p_1 \in [0,1]$ and $p_\infty = 0$, and $M$ where both $p_1 \in [0,1]$ and $p_\infty \in [0,1]$. Hence, any difference between results obtained under $M_0$ and $M$ reflect the impact of subsequent exposure, with $p_\infty$ controlling the magnitude of the effect. We deploy both model variants $M$ and $M_0$ to a real-world network of activities (i.e. a *project* [33,34]), and explore two key quantities that characterise spreading – the spreading event size ($S_n$) and the rate by which spreading propagates ($S_r$) –where $p_\infty$ has an important effect. We subsequently focus on particular structural features of the pathways used to sustain these spreading events, and specifically on the different impact that $p_1$ and $p_\infty$ have. Finally, we explore the topological properties of nodes involved in large-scale spreading events. In agreement with recent studies[6,35] on information spreads, we report the presence of 'hidden influentials' (i.e. nodes with average topological properties which play a key role in sustaining large-scale spreading events), with their existence being increasingly pronounced at higher $p_1$ and/or $p_\infty$ values.

**Results**

We first establish the importance of subsequent exposure (parameter $p_\infty$) by illustrating its effect on the spreading event size through a comparative analysis between the results of spreading model $M$ and $M_0$. At this point we provide evidence on the link between subsequent exposure and clustering, where an increase in the latter (clustering) controls the magnitude of the former (subsequent exposure). We subsequently use model $M$ to highlight the contrasting impact that direct (parameter $p_1$) and subsequent exposures have on propagation rate. We then focus on the relationships between the structural characteristics of these spreading events and their size/rate, and how $p_1$ and $p_\infty$ affect them. Finally, we provide insight on the topological features of nodes capable of fuelling large-scale spreading events, and how $p_1$ and $p_\infty$ influence them.

*Spreading event size and propagation rate*

We first define the ratio of the largest spreading event size obtained using the $M$ model, over the largest spreading event size obtained under the $M_0$ model, as $r^{\max} = \frac{M(S_n^{\max})}{M_0(S_n^{\max})}$. In a similar fashion, we denote the ratio of the average spreading event sizes as $r^{\text{avg}}$. With $r^{\max}$ and $r^{\text{avg}}$ being functions of both $p_1$ and $p_\infty$, we can examine their isolated effect by considering their corresponding averages: $\tilde{r}^{\max}(p_1) = \frac{1}{|p_\infty|}\sum_{p_\infty=0}^{1} r^{\max}(p_1, p_\infty)$ and $\tilde{r}^{\max}(p_\infty) = \frac{1}{|p_1|}\sum_{p_1=0}^{1} r^{\max}(p_1, p_\infty)$. By applying the same approach to $r^{\text{avg}}$, we obtain $\tilde{r}^{\text{avg}}(p_1)$ and $\tilde{r}^{\text{avg}}(p_\infty)$.

Increasing parameter $p_\infty$ leads to values of $r^{\max}$ (Fig. 2a) and $r^{\text{avg}}$ (Fig. 2b) being significantly higher than 1, demonstrating the augmenting effect that subsequent exposures have on the spreading event size. This result suggests that this activity network contains a high enough number of non-trivial subgraphs – such as the ones included in Fig. 1 – which allows $p_\infty$ to have a significant impact on the progression of the spreading process. If the converse where to be true (i.e. $r^{\max} \cong r^{\text{avg}} \cong 1$), it would suggest that the activity network could be approximated by 'locally tree-like' network, where subsequent exposure would have had no effect over the spreading process, with results resembling that of a tree network (results from a tree network are shown in Supplementary Fig. 2 for reference).

Taken in isolation, $p_1$ and $p_\infty$ show qualitatively different characteristics in terms of their impact in the spreading process, further highlighting the non-trivial interaction between direct and subsequent exposure in the context of a spreading process. The concave relationship between both $\tilde{r}^{\max}$ and $\tilde{r}^{\text{avg}}$, with respect to $p_1$, demonstrates the principal role of $p_1$ in sustaining the spreading process, with both $\tilde{r}^{\max}$ and $\tilde{r}^{\text{avg}}$ converging to 1 at the two extreme ends of $p_1$ (Fig. 2c). On one hand, if $p_1 = 0$ no spreading occurs and therefore the effect of $p_\infty$ is nullified, with both $M$ and $M_0$ converging to identical spreading events; when $p_1 = 1$ then direct exposure successfully switches the state of *all* nodes and therefore, no nodes are left for $p_\infty$ to affect. Interestingly, high $\tilde{r}^{\max}$ values are preserved up to relatively high $p_1$ values ($p_1 \leq 0.7$), indicating the strong influence of $p_\infty$ even under unfavourable conditions i.e. under $p_1 = 0.7$, a node is much more likely to switch state due to a direct rather than subsequent exposure, and therefore one would naturally expect that the influence of $p_\infty$ would be limited, giving rise to a low $\tilde{r}^{\max}$ value. Finally, the intuitive expectation of $p_\infty$ having an ever-increasing effect in terms of both $\tilde{r}^{\max}$ and $\tilde{r}^{\text{avg}}$ is supported by the monotonically increasing trends shown in Fig. 2d.

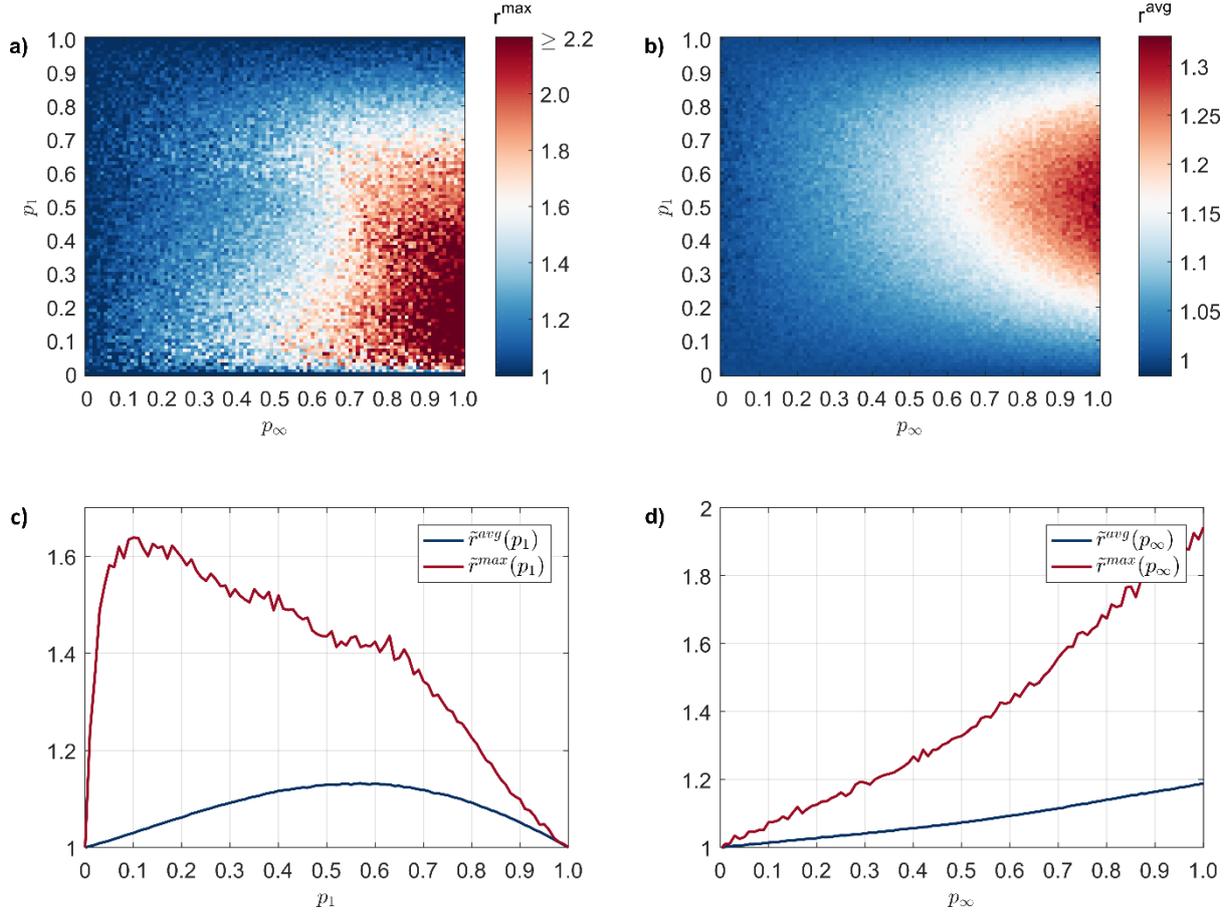

**Fig. 2**: Difference in spreading event size between spreading model $M_0$ and $M$ (a) Ratio of the largest spreading event sizes ($r^{\max}$) under the entire spectrum of $p_1$ an $p_\infty$ (largest $r^{\max} = 3.2$); (b) like (a), focusing on the ratio of average spreading event size ($r^{\text{avg}}$) under parameter $p_1$ and $p_\infty$; (c) ratio of largest (red) and average (blue) spreading event size, averaged across $p_\infty$, as a function of $p_1$; (d) ratio of largest (red) and average (blue) spreading event size, averaged across $p_1$, as a function of $p_\infty$.

This insight is consistent when we consider the cumulative probability distribution of $S_n$, focusing on (i) the probability of observing a spreading event of a given size, and (ii) the magnitude of the largest event. In this case, the augmenting role of $p_\infty$ is particularly pronounced at the tail of the distribution, where higher $p_\infty$ increases both (i) and (ii) (Fig. 3b). This is in contrast to the direct impact of $p_1$, where highger $p_1$ increase both(i) and (ii) across the entire range of $S_n$ (Fig. 3a). Taken in conjunction, this behaviour demonstrates the subtle impact of indirect interactions with respect to the emergence of small-scale spreading events – which are largely driven by direct interactions – and their marked influence with respect to large-scale spreading events, both in terms of their probability of their emergence and their absolute size.

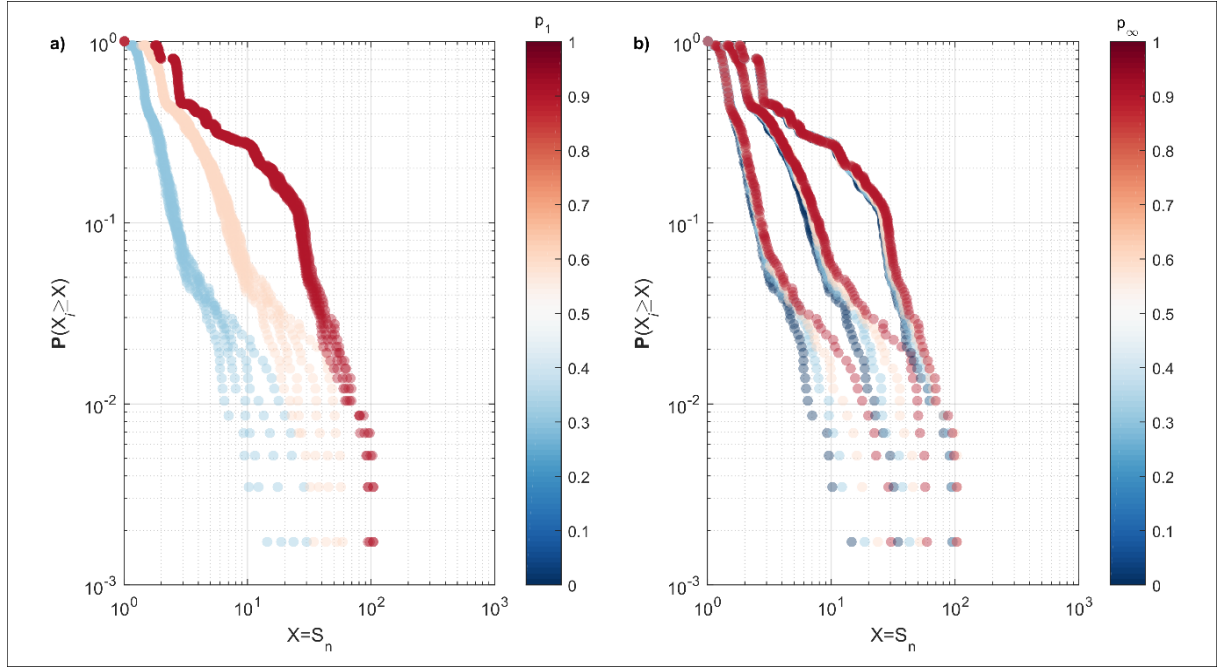

**Fig. 3**: Cumulative probability distribution of spreading event size ($S_n$), where marker colour corresponds to (a) $p_1$ and (b) $p_\infty$. For clarity, $p_1$ and $p_\infty$ are sampled in 0.3 step intervals, taking the values of $[0, 0.3, 0.6, 0.9]$.

The extent by which subsequent exposures control the spreading event size depends on the clustering of the network, as captured by the clustering coefficient ($C$) [36]. To demonstrate this, we have deployed the Watts-Strogatz model to generate artificial networks whilst varying the rewiring probability $\beta$, moving progressively from clustered ($\beta = 0.1, C = 0.3$) to random networks ($\beta = 1, C = 0.03$). In doing so, we find that an increase in $\beta$ (and hence, drop in $C$) decreases the difference between spreading event sizes obtained by $M$ and $M_0$, for both $r^{\text{avg}}$ and $r^{\text{max}}$, with results being qualitatively similar to Fig. 2 (see Fig. 4). This finding suggests that the increased impact that subsequent exposure has to the spreading process in general – and to the activity network in particular – relies, at least partly, to increased clustering.

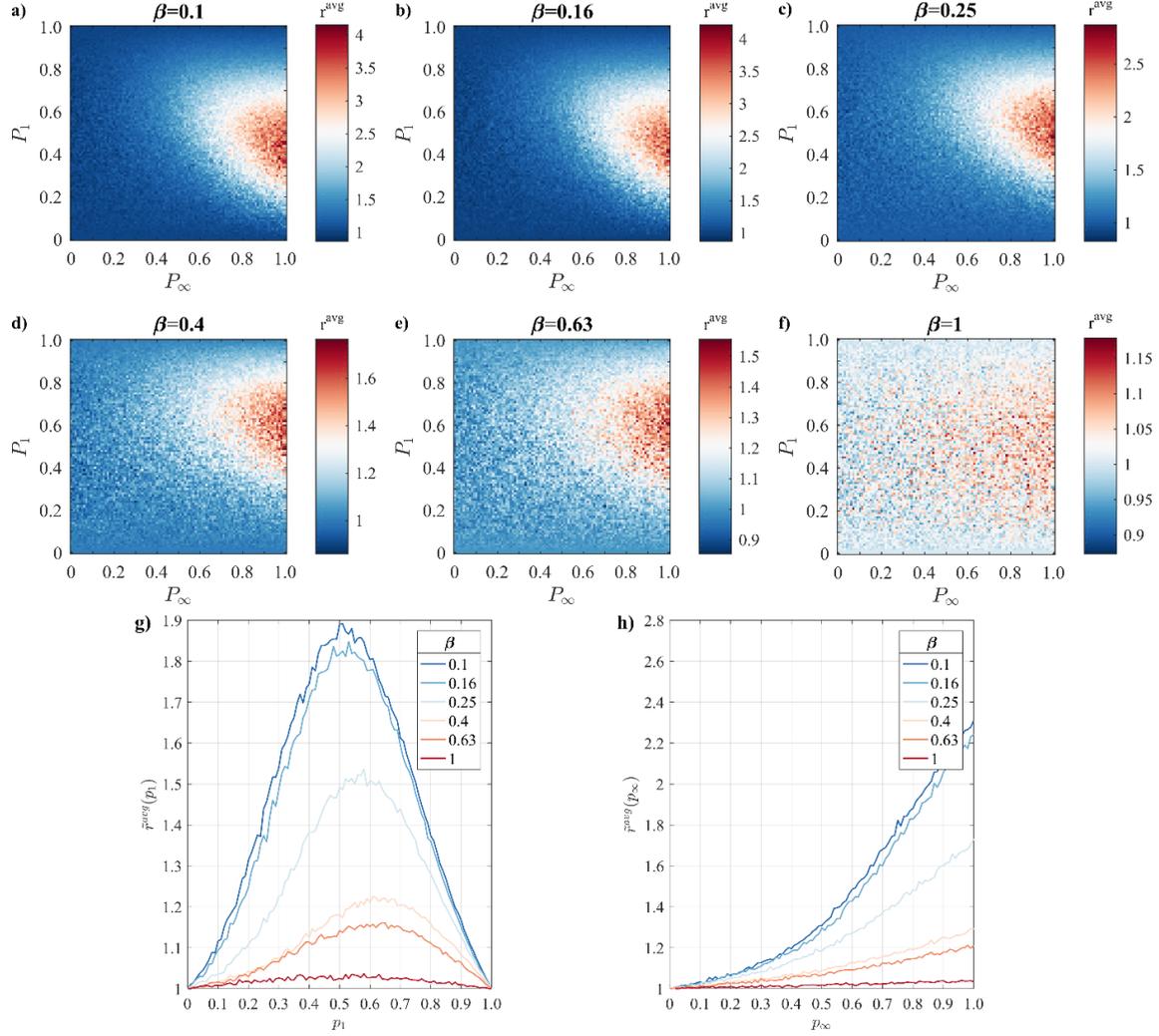

**Fig. 4**: Difference in spreading event size between spreading model $M_0$ and $M$ for a set of artificial, 'small-world' networks with increasingly probability of rewiring ($\beta$), ranging from increasingly clustered (a; $\beta=0.1$) to random (f; $\beta=1$) topologies; (g) ratio of average (blue) spreading event size, averaged across $p_\infty$, as a function of $p_1$, across the $\beta$ spectrum; (h) ratio of average (blue) spreading event size, averaged across $p_1$, as a function of $p_\infty$, across the $\beta$ spectrum.

We now focus on the rate by which a given spreading event propagates across the network, which we quantify as $S_r = \frac{S_n}{t}$, where $S_n$ refers to the spreading event size and $t$ refers to the average number of simulation steps needed for all nodes affected (within that spreading event) to switch state from 'non-affected' to 'affected' (which is a variation of survival probability [37]). As such, we define the average propagation rate, $\tilde{S}_r^{avg}$ as the propagation rate for each spreading event, averaged across all events, and the maximum propagation rate, $\tilde{S}_r^{max}$, as the propagation rate for the single largest spreading event.

We find that direct and subsequent exposures have the converse effect with respect to the propagation rate, both in terms of $\tilde{S}_r^{avg}$ and $\tilde{S}_r^{max}$. In particular, we find a positive relationship between $p_1$ and the propagation rate, in terms of both $\tilde{S}_r^{avg}$ (blue marker) and $\tilde{S}_r^{max}$ (red marker), which corresponds to the intuitive expectation where increased direct exposure eases the way in which spreading progresses, enhancing the overall propagation rate (Fig. 5a). However, a negative relationship exists between $p_\infty$ and propagation rate, both in terms of $\tilde{S}_r^{avg}$ and $\tilde{S}_r^{max}$ (Fig. 5b). This is due to the elaborate topology of the pathways deployed by the spreading process, where higher $p_\infty$ increases the likelihood of utilising wider spreading pathways (i.e. high $S_w$; see *Topology of spreading pathways*) which are more likely to involve a higher number of links to be traversed for affecting the same number of nodes (increasing the denominator of $S_r$) eventually delaying the overall spreading process. Considering both effects, this behaviour suggests that propagation rate is conflated by contrasting dynamics, where direct exposure provides immediate – and thus, faster – pathways for spreading to propagate, while subsequent exposure unlocks slower pathways which supress the overall in propagation rate (even though they may increase the overall spreading event size, as seen in Fig. 2d).

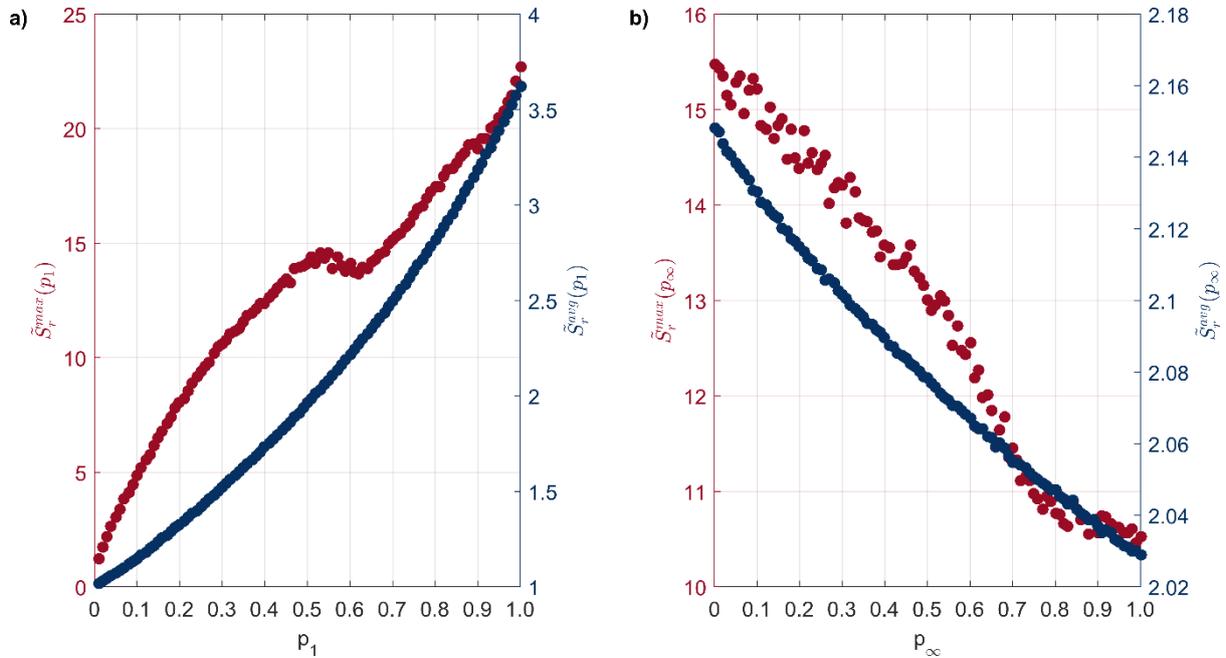

**Fig. 5**: Propagation rate as a function of (a) direct exposure, $p_1$, and (b) subsequent exposure, $p_\infty$. Blue and red markers correspond to the average rate across all spreading events, $\tilde{S}_r^{\text{avg}}$, and propagation rate for the largest spreading event, $\tilde{S}_r^{\max}$, respectively.

*Topology of spreading pathways*

We characterise the structure of a spreading event by considering the maximum depth and width of the underlying pathways that have sustained it. We define the maximum depth of a spreading event ($S_d$) as the shortest path between the initial seed node and the farthest node involved in the event [6]. In addition, we define its maximum width ($S_w$) as the maximum number of nodes affected whilst being at the same distance from the seed node [38]. As such, we characterise the structure of each spreading event as $S_{d/w} = \frac{S_d}{S_w}$, which provides a continuous measure for the overall shape of the underlying pathways, where a high value of $S_{d/w}$ corresponds to long and narrow pathways, whilst a low value of $S_{d/w}$ corresponds to short and wide pathways (see Supplementary Fig. 3 for examples). In that way, $S_{d/w}$ is maximised when the spread is composed of a single linear chain, and minimised when the spread resembles the structure of a star-shaped network.

We first focus on the relationship between the largest spreading event size, $S_n^{\max}$ as a function of $S_{d/w}$, where we identify a non-trivial relationship roughly composed of two opposing trends, see Fig. 6 ($S_n^{\max}$ is normalised over the total number of nodes, $N$). The first trend dominates the small to medium sized events, where the spreading event size increases in step with $S_{d/w}$, demonstrating the reliance of the spreading process to long and narrow pathways. However, as spreading events become larger than a given threshold (in this case, when $\frac{S_n^{\max}}{N} \geq \approx 0.05$) the positive relationship between $\frac{S_n^{\max}}{N}$ and $S_{d/w}$ reverses, with the spreading process enlisting an increasingly high number of relatively shorter and wider pathways. This switch suggests the existence of an upper bound in the number of long and narrow sequence of consecutive tasks within the activity network. These sequences are largely composed of low-out degree nodes, and given the finite size of the network, pose a limit to the growth of the spread. To surpass this limit, and to further fuel the growth of the spreading event size, spreading utilises additional pathways which emerge through the inclusion of occasional high out-degree nodes, which allow for the spreading process to branch out in order to increase in size, resulting in relatively wider spreads.

The rate by which the spreading event size increases depends on the spreading event structure. In the case where the spreading event size is negatively correlated with $S_{d/w}$, the rate by which the spreading event size grows is roughly 3 times faster compared to the case where the spreading event size is positively correlated with $S_{d/w}$ (gradient is roughly -0.09 and +0.03, respectively). This result emphasizes the multiplicative effect that wider structures can provide, which in turn enhances the number of nodes that can be reached, and in turn, affected.

With respect to the impact of direct and subsequent exposure, $p_1$ and $p_\infty$ show district trends, demonstrated by the marker colour patterns in Fig. 6a and 6b, respectively. Focusing on the impact of $p_1$, the transition in marker colour, from blue to red, is accompanied with a smooth increase in the spreading event size (Fig. 6a, 6c). This result is somewhat expected, since $p_1$ plays a key role in the progression of the overall spreading process. In addition, $p_1$ has an important role in determining the structure of the resulting spreading structure, albeit in a non-trivial manner. In particular, short and wide structures (low $S_{d/w}$) can occur at both extremes of $p_1$, with the shape slowly converging to the highest attainable $S_{d/w}$ values as $p_1$ approaches ~0.5. Shifting focus to the impact of $p_\infty$, we observe that the entire range of $S_{d/w}$ is obtainable under any given value of $p_\infty$, indicating the limited role of subsequent exposure in determining the structure of the pathways used by the spreading process (Fig. 6b). The subtle impact of $p_\infty$ on $S_{d/w}$ is further highlighted in the limited range of $S_{d/w}$ obtained reported Fig. 6d, which is significantly lower than the corresponding impact of $p_1$ in Fig. 6c.

Note that these results are robust when we consider the relationship between $S_n^{\text{avg}}$ (instead of $S_n^{\text{max}}$) and $S_{d/w}$, as a function of $p_1$ or $p_\infty$, see Supplementary Fig. 4.

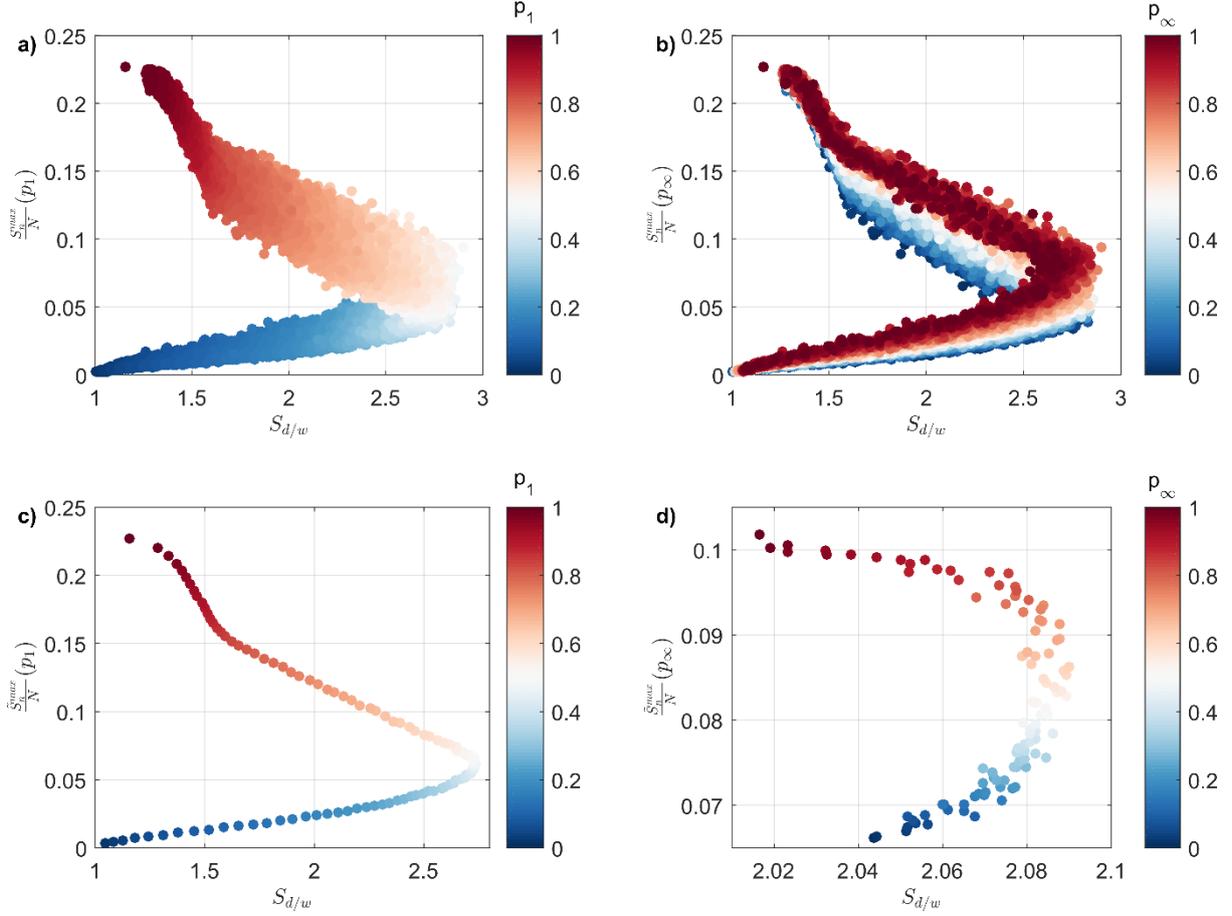

**Fig. 6:** Parameter space of the largest spreading event size ($S_n^{\max}$), normalised over the total number of nodes ($N$), and its underlying structure ($S_{d/w}$), as a function of (a) $p_1$ and (b) $p_\infty$; (c) the relationship between the largest spreading events, averaged over $p_\infty$ and mapped as a function of $p_1$, and (d) the relationship between the largest spreading events, averaged over $p_1$ and mapped as a function of $p_\infty$.

We now focus on the relationship between the propagation rate of the largest spreading event, $S_r^{\max}$, and the structure of the pathways used to sustain it, $S_{d/w}$, as a function of $p_1$ (Fig. 7a) and $p_\infty$ (Fig. 7b). Similar to Fig. 7, we identify a non-trivial relationship roughly composed of two distinct behaviours, where the propagation rate initially increases in step with spreading pathways becoming increasingly long and narrow (higher $S_{d/w}$). This overall increase in $S_r^{\max}$ is the result of two conflicting effects – an increase in propagation rate, driven by higher $p_1$ values (Fig. 7a and 7c), combined with a decrease in propagation rate driven by higher $p_\infty$ values (Fig. 7b and 7d). This result demonstrates the conflicting nature of $p_1$ and $p_\infty$, where the former relies on direct interactions which are faster to affect, while the latter introduces additional indirect pathways that take longer to evaluate completely due to their non-trivial nature (similar to the ones depicted in Fig. 1).

Once $S_r^{\max}$ reaches a given threshold (in this case, $S_r^{\max} \approx 12$), its positive relationship with $S_{d/w}$, reverses to a negative relationship, which essentially reflects the need to utilise wider pathways (and hence, triggers a decrease in $S_{d/w}$), in order to increase the propagation rate further. This switch in behaviour is induced when $p_1$ grows over ~0.5; as soon as this reversal in the relationship between $S_r^{\max}$ and $S_{d/w}$ takes place, the fork-like shape of Fig. 7 indicates that two possible trajectories are available. Importantly, the similar marker colouring within both trajectories (Fig. 7a) suggests that $p_1$ has a limited role in determining which of the two trajectories is followed. Yet the distinct marker colouring shown in Fig. 7b indicates that $p_\infty$ is the key parameter in determining which of the two trajectories is followed. Specifically, the primary trajectory is accessible under the entire range of $p_\infty$, and the secondary trajectory, accessible under a limited range of $p_\infty$ values, roughly ranging from 0 to 0.6. This demonstrates the complex nature of subsequent exposure: on one hand, low-to-medium subsequent exposure means that the topology of the utilised failure pathways depends on the topology of the network itself; on the other hand high subsequent exposure always results in wider pathways being utilised (since any auxiliary path stemming from the main failure pathway is pursued to its full length, triggering an increase in $S_w$, and hence a decrease in $S_{d/w}$). Both of these aspects are further reinforced by isolating the results obtained at equal $p_1$ and $p_\infty$ increments, with $p_\infty$ controlling the emergence of the second trajectory (see Supplementary Fig. 5 for $p_1$, and Supplementary Fig. 6 for $p_\infty$).

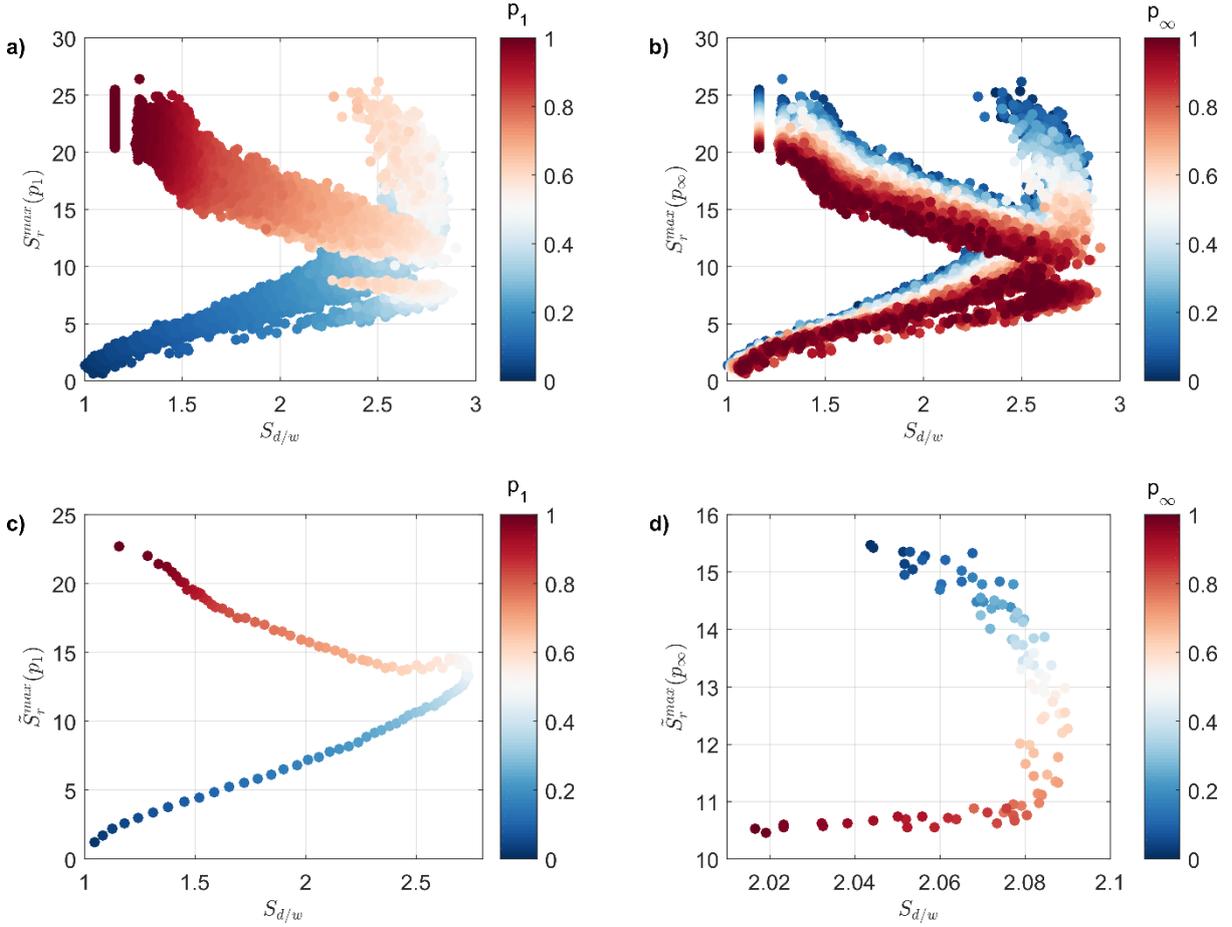

**Fig. 7:** The relationship between the propagation rate of the largest spreading events across the entire range of $p_1$ and $p_\infty$, $(S_r^{max})$, and its underlying structure $(S_{d/w})$, as a function of (a) $p_1$ and (b) $p_\infty$; (c) the relationship between the propagation rate of the largest spreading events, averaged over $p_\infty$ and mapped as a function of $p_1$, and (d) the relationship between the propagation rate of the largest spreading events, averaged over $p_1$, and mapped as a function of $p_\infty$.

*'Hidden influentials' and the effect of $p_1$ and $p_\infty$*

Large-scale spreading events are typically associated with extra-ordinary topological characteristics of the node that initially triggers them (i.e. the seed node), the simplest of which being the (out) degree[39] – our results confirm this common conjecture, albeit with certain strong caveats. Specifically, we find that the spreading event size is highly correlated with node out-degree, as reported in [6,40] (Fig. 8a). However, this correlation deteriorates as $p_1$ increases in size, which suggests that the spreading process is shifting from being a local-driven process (and hence, dominated by the properties of the seed node) to a globally-driven process, where the characteristics of the intermediate nodes eventually dilute the correlation between spreading event size and the topological characteristics of the seed node. This correlation deteriorates faster once the effect of $p_\infty$ is introduced, as additional intermediate nodes are employed early on during the spreading. Recent empirical work on information spreads has identified a similar effect, where large-scale spreading events are largely sustained by intermediate nodes with no special topological features – the so-called 'hidden influentials'[6,41].

Following the work of Baños, et al. [6], we evaluate whether these 'hidden influentials' exist in the activity network by comparing the average out-degree of nodes involved in a spreading event, excluding that of the seed node, $\left(\tilde{k}_{\text{out}}^{\text{avg}}\right)$ with the network average $\left(k_{\text{out}}^{\text{avg}}\right)$, and the relationship between $\tilde{k}_{\text{out}}^{\text{avg}}$ and the spreading event size. Notably, $\tilde{k}_{\text{out}}^{\text{avg}}$ converges to $k_{\text{out}}^{\text{avg}}$ as the spreading event size increases, confirming the presence of these 'hidden influentials' (Fig. 8b). This behaviour demonstrates that the existence of extra-ordinary nodes (e.g. hubs) is not a necessary condition for large-scale spreading to occur, and hints for other non-local topological properties that may characterise the nature of these 'hidden influentials' [9,42]. More generally, this result highlights the intrinsic challenge in containing spreading in general, where system-wide spreading events are sustained by merely typical nodes, which themselves are hard to identify *a priori*.

We now focus on exploring the impact of $p_1$ and $p_\infty$ on the emergence of these 'hidden influentials', by considering the largest $\tilde{k}_{\text{out}}^{\text{avg}}$, averaged across the entire $p_\infty$ and $p_1$ values, respectively. Notably, an overall decreasing trend is noted as both $p_1$ and $p_\infty$ increase, where an increase in $p_1$ triggers a rapid decrease in $\tilde{k}_{\text{out}}^{\text{avg}}$ (Fig. 8c), converging towards $k_{\text{out}}^{\text{avg}}$, while an increase in $p_\infty$ triggers a linear decrease in $\tilde{k}_{\text{out}}^{\text{avg}}$ (Fig. 8d). This behaviour corresponds to an increase in the role of 'hidden influential' in sustaining the spreading process as immediate failure becomes more likely. In conjunction with the fact that spreading event sizes increases with larger $p_1$ and/or $p_\infty$, these results further suggest that larger spreading events may be harder to contain than smaller ones, simply because larger ones are increasingly reliant on the existence of these 'hidden influentials'.

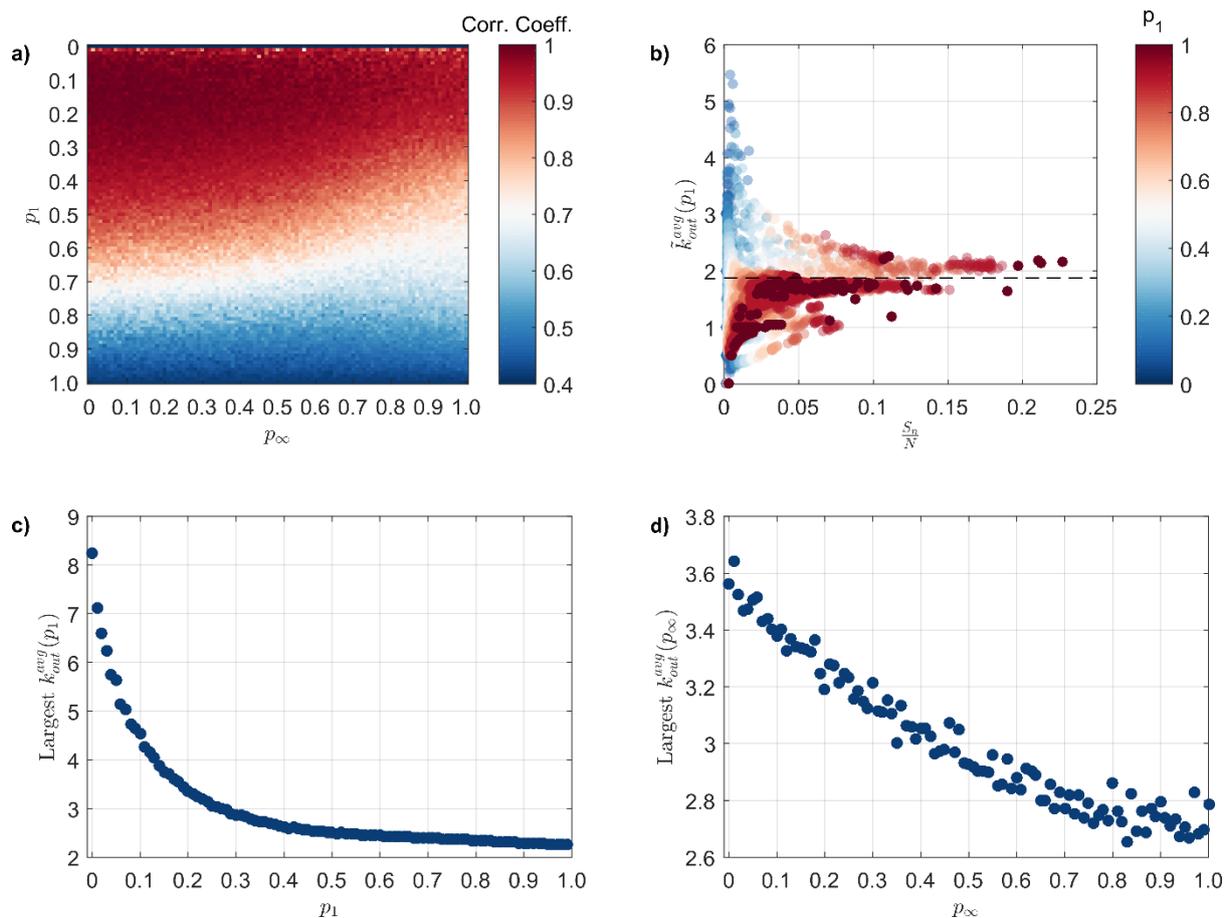

**Fig. 8**: (a) Correlation coefficient between spreading event size and the out-degree of the initial seed node, as a function of $p_1$ and $p_\infty$; (b) average out-degree of nodes involved in a spreading event, excluding that of the initial seed node, $\left(\tilde{k}_{out}^{avg}\right)$ as a function of the normalised spreading event size and $p_1$; dotted line corresponds to the network average out-degree, $k_{out}^{avg}$; (c) largest $\tilde{k}_{out}^{avg}$ as a function of $p_1$, where a lower value indicates an increasingly important role for typical node in sustaining spreading, converging to $k_{out}^{avg}$; (d) largest $\tilde{k}_{out}^{avg}$ as a function of $p_\infty$

**Discussion**

In this paper, we have introduced a simple model which allows us to decouple the effect of subsequent exposure from the overall spreading process, and comparatively examine its impact on key quantities, including spreading event size (Fig. 2,3) and propagation rate (Fig. 5). Our results highlight the conflating nature of spreading, where subsequent exposure increases the number of nodes affected whilst reducing the rate in which the spread progresses. With subsequent exposure being a derivative of clustered networks, our results clarify broader discussions within the literature, which typically focus in providing high-level insight[43-46]. For example, ref [44] focuses on identifying a (largely) positive link between clustering and spreading event size. Yet, the question of *why* clustering enhances spreading event size remains unexplored, with cases of no effect being treated as some sort of outliers. Our results suggest that subsequent exposure is one possible avenue by which the relationship between clustering and enhanced spreading event size depends on, allowing for additional aspects to be explored in a similar fashion.

From a methodological standpoint, our results demonstrate the need to explicitly account for, and control the effects of, subsequent exposure when modelling spreading-like processes. For example, consider the frequent use of 'locally tree-like' approximations, typically used to deploy analytically tractable expositions into various network dynamics[1,47-49]. Despite the valuable insights that these approximations provide, the eventual nullification of subsequent exposure – and its effect on the spreading process – clouds the real difference between these models and the respective real-world systems they represent, skewing our confidence and biasing results in a non-trivial manner. The results presented within this paper serve as additional motivation to recently emerging lines of inquiry[50,51] which focus on relaxing the 'locally-tree like assumption', integrating the effect of subsequent exposure to the overall spreading process. More generally, relaxing these approximations has the potential for uncovering dynamical properties that are shared across a range of real-world systems, similar in spirit to the work of Barzel and Barabási [20].

In terms of applications, our work provides the grounds for a dialogue between researchers in the network science and project management, where hotly-researched, domain challenges (e.g. project complexity evaluation[52-57]) can be treated as network-related problems [31,32]. For example, increased susceptibility to the spreading of failures can be reasonably interpreted as a contributing factor to project complexity. Hence, the relationship between spreading event size (or propagation rate) and the structure of the underlying pathways (Fig. 6 and 7 respectively) can serve as an objective, quantitative measure for project complexity (assuming that the activity network is an up-to-date reflection of the actual project plan). Similarly, the proportion (and identity) of 'hidden influentials' within an activity network (Fig. 8) could be used to support the overall project risk mitigation scheme, where activities with limited connections (yet increased probability of being 'hidden influentials') receive adequate attention.

**Methods**

*Data*

The data comprises of a real-world engineering project which captures a set of planned activities that need to be completed in order to deliver a definitive commercial product in the area of defence. The overall duration of the project is 577 days, and is composed of 578 distinct tasks with 1,085 dependencies. Note that some tasks are used as planning instruments (e.g. milestones [58]) and thus, include no dependencies – these tasks are excluded from the analysis (8 tasks in total).

The delivery of each activity is typically conditional to a number of other activities. We refer to these interactions as functional dependencies, since they effectively control the function of each activity e.g. the start of activity $j$ depends on the completion of activity $i$. The directionality of each dependency dictates the functional role of each activity i.e. whether it acts as a predecessor (activity $i$ proceeds activity $j$) or a successor (activity $k$ succeeds activity $j$) to a subsequent task (leaf activities also exist, with no successor activities).

The set of tasks and dependencies was subsequently converted to an activity network, defined as a directed network $G = \{V, E\}$, where $V$ is the set of nodes and $E$ is that of directed edges. Every activity is abstracted as a node, where a functional dependency between activity $i$ and $j$ is captured in the form of a directed link from node $i$ to node $j$, denoted by $e_{ij} \in E$. The number of successors and predecessors each activity has corresponds to its out-degree and in-degree respectively. The cumulative probability distribution of out-degree (red) and in-degree (blue) is shown in Supplementary Fig. 7– note its heavy-tail nature, evident by the straight line formed under the log-log axes.

*Spreading Model Formulation*

Every node $j$ of the network at time $t$ is characterised by a dynamic variable $s_j(t) \in \{0,1\}$, where '0' and '1' correspond to the 'non-affected' and 'affected' state, respectively. During the spreading process, node $j$ may irreversibly switch from the 'non-affected' to the 'affected' state at time $t$ if: (i) node $j$ has at least one predecessor, node $i$, and (ii) at least one node $i$ was in the 'affected' state at $t -$

1. Then, we artificially switch the state of some seed node at $t = 0$, from 'non-affected' to 'affected' and track the progression of the spreading process as time increases at discrete increments of 1.

In order to distinguish between direct and subsequent exposures, we keep track of node $i$'s successors at time $t$, say node $j$, and assess whether node $j$ has been encountered before. If so, then node $j$ has been directly exposed to the failure of a *different* predecessor at some time $> t$, but did not switch states during that time. Therefore, the probability of node $j$ to switch states at now, at time $t$, is controlled by $p_\infty$ (Fig. 1, middle panel). However, if this is the first time node $j$ has been encountered, then it is the first time node $j$ is exposed to failure in general and therefore the probability to switch states at time $t$ is controlled by $p_1$ (e.g. Fig. 1, top panel).

Note the broad nature of the term 'affected', acknowledging the fact that failure can mean very different things, depending on the context of the project. For example, 'failure' can mean 'structural defect' in a construction project, or something much less tangible such as a 'contaminated' or 'compromised' in a cyber-security project.

*Spreading Model Implementation*

The model is implemented as follows. First, an initialisation phase is implemented, where simulation time $t$ is set to 0 and the state of all nodes is set to '0'. In addition, an empty set $B$ is created in order to record all successor nodes encountered during time $t$. The spreading process is initiated by externally switching the state of node $i$ from '0' to '1'. We then identify all successors of node $i$, node(s) $j$ (if no neighbours exist, the process terminates). For each node $j$, we record index $j$ in set $B(t)$, and then check whether index $j$ was already present in set $B(t-1)$. If index $j$ was not present, the interaction between node $i$ and $j$ is the result of direct exposure; hence, the probability of node $j$ to switch states, under both model $M$ and $M_0$, is equal to $p_1$. However, if index $j$ was already present in set $B(t-1)$, the interaction between node $i$ and $j$ is the result of subsequent exposure; hence, the probability of node $j$ to switch states, under model $M$, is equal to $p_\infty$ (in the case of $M_0$, $p_\infty$ is always set to 0). Once all node(s) $j$ have been tested with respect to the prospect of changing states, we record the total number of state changes up to, and including, time $t$, $S_n(t)$, and increase $t$ by 1. The process repeats until the total number of state changes remains constant i.e. $S_n(t) = S_n(t-1)$. Finally, the process is reiterated for each node $i$, in order to evaluate the total number of state changes the failure of every possible seed node. Finally, this process is repeated for 48 independent runs, with results presented herein being the average (number of runs determined in order to minimise the standard error of the mean).

Formally, the condition by which node $j$ changes state depends on the spreading model used. The condition for spreading model $M_0$ is determined by eq.1:

$$s_j(t) = \begin{cases} 1, & \text{if } j \notin B(t-1) \land P_1 \geq \Theta, \quad P_1 \in [0,1] \\ s_j(t-1) & \text{otherwise} \end{cases} \quad \text{(eq.1)}$$

and the condition for spreading model $M$ is determined by eq.2:

$$s_j(t) = \begin{cases} 1, & \text{if } j \notin B(t-1) \land P_1 \geq \Theta, \quad P_1 \in [0,1] \\ 1, & \text{if } i \in B(t-1) \land P_\infty \geq \Theta, \quad P_\infty \in [0,1] \\ s_j(t-1) & \text{otherwise} \end{cases} \quad \text{(eq.2)}$$

where variable $\Theta$ is uniformly drawn at random from $U(0,1)$ for both eq.1 and eq.2.

*Data availability*

The datasets generated and/or analysed during the current study, and related source code, are available from the corresponding author on reasonable request.

**Acknowledgements**

CE was partly funded by Thales and an EPSRC Doctoral Prize fellowship. CE is particularly grateful to Angus Johnson of Thales UK for providing the dataset, and Naoki Masuda for insightful discussions.

**Additional Information**

*Author contributions*

CE obtained funding, designed the research, performed numerical simulations, analysed the results and prepared the manuscript.

*Competing financial interests*

The author(s) declare no competing financial interests.

# Supplementary Material for 'Modelling indirect interactions during failure spreading in a project activity network'

Christos Ellinas[1,*]

[1] *Engineering Mathematics, University of Bristol, Bristol, UK*

[*] Corresponding author

E-mail: ce12183@bristol.ac.uk

## 1. Visualisation of empirical activity network

Supplementary Fig. 1 visualises the empirical activity network, with subplot (b) focusing on an exemplar spreading event which includes several instances of non-tree-like behaviour – this behaviour is highlighted by grey nodes. Notice the resemblance of the highlighted topology with the motivational example used in Fig.1 of the main paper.

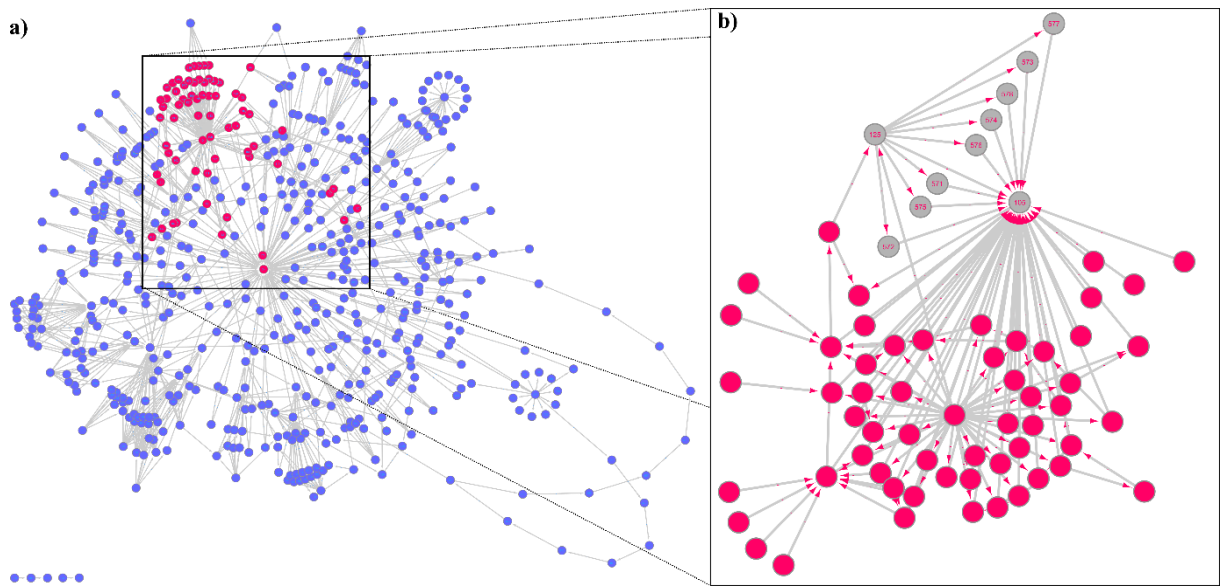

**Supplementary Fig. 1:** Visualisation of (a) the empirical activity network, highlighting an exemplar spreading event (red nodes), (b) exemplar spreading event, highlighting the deviation from tree-like behaviour (grey nodes). Note the resemblance to the example used in Fig. 1.

## 2. Tree network case

The average spreading event size obtained under model $M_0$ (Supplementary Fig. 2a) is visually indistinguishable from those obtained under model $M$ (Supplementary Fig. 2b). This is increasingly evident once we consider $r^{\text{avg}}$ (Supplementary Fig. 2c), the results of which converge to a value of 1, with some normally distributed noise that results from the stochastic nature of the simulation (Supplementary Fig. 2d). This coherence between model $M_0$ and $M$ illustrates the lack of an effect from indirect exposures, which is to be expected, since non-trivial subgraphs such as the ones noted in Figure 1, as they are explicitly forbidden by the very structure of the tree network.

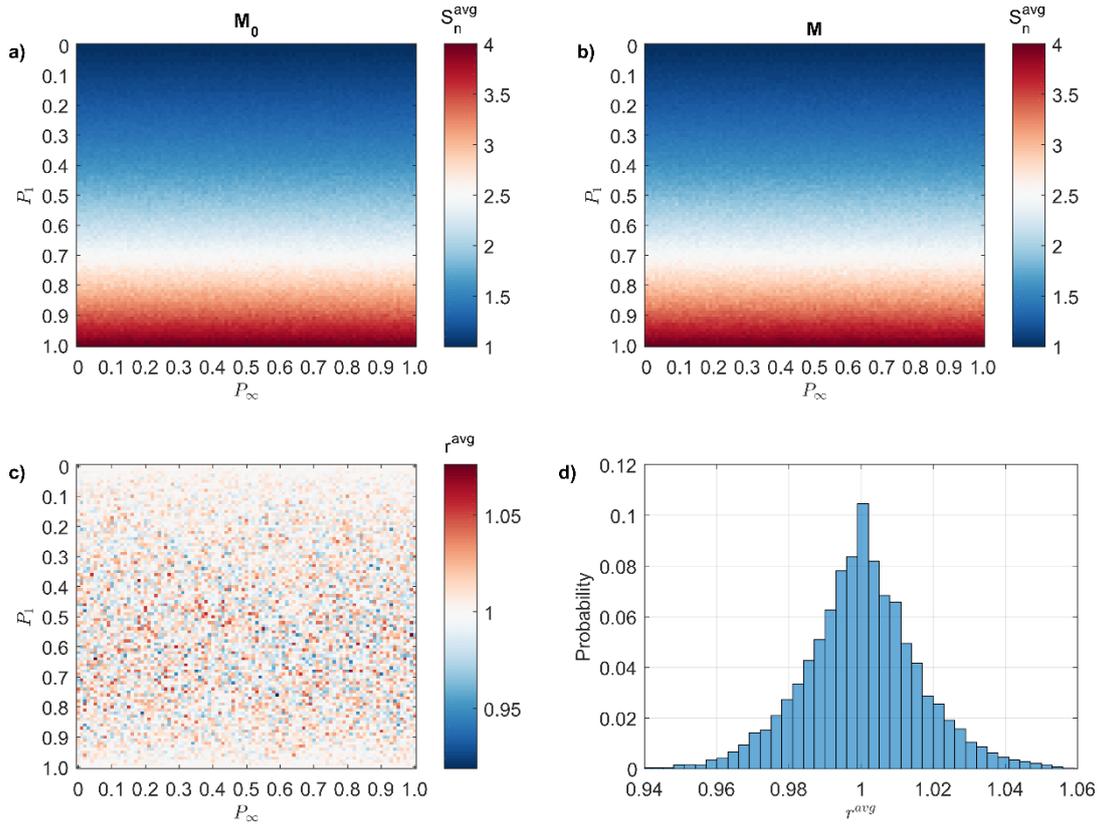

**Supplementary Fig. 2:** (a) spreading event size across the parameter space under model $M_0$; (b) same as (a) under model $M$; (c) value $r^{avg}$ across the parameter space, demonstrating the lack of an effect as indirect exposure increases; (d) histogram of $r^{avg}$ values in (c), converging to a value of 1 (no difference between $M_0$ and $M$) with some normally distributed noise due to the stochastic nature of the simulation.

## 3. Example subgraphs for varying $S_d$ and $S_w$

Supplementary Fig. 3 provides simple examples for varying $S_d$ and $S_w$ used to characterise the topology of the spreading pathways, whilst preserving the size of the spreading event ($S_n$) constant (in this case, $S_n = 5$). Nodes that contribute to the final $S_d$ and $S_w$ are marked using dotted and blue borders, respectively

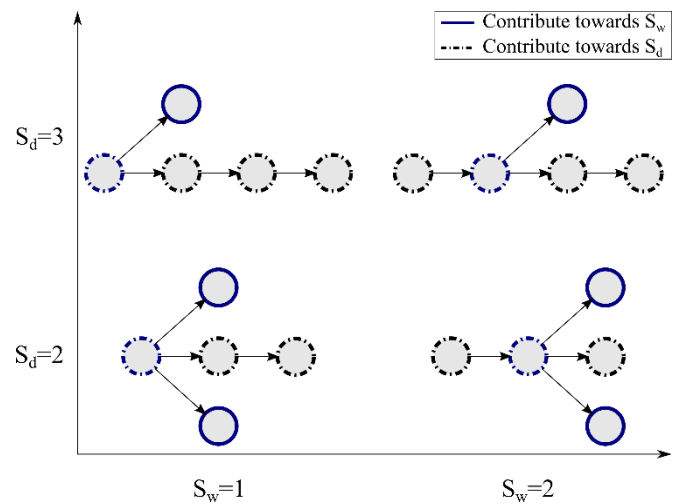

**Supplementary Fig. 3:** A variety of spreading pathways that capture the failure of six nodes, as characterised by varying $S_d$ and $S_w$.

## 4. Average Spreading Event Size

Similar to the case of $S_n^{\max}$ reported in the main manuscript (Figure 6), a similar relationship between $S_n^{\max}$ and $S_{d/w}$ is noted, where small to medium events rely on narrow pathways (high value of $S_{d/w}$), see Supplementary Fig. 4a and 4b respectively. Larger events are subsequently enabled by branching out to include additional pathways which result to wider pathways (low value of $S_{d/w}$).

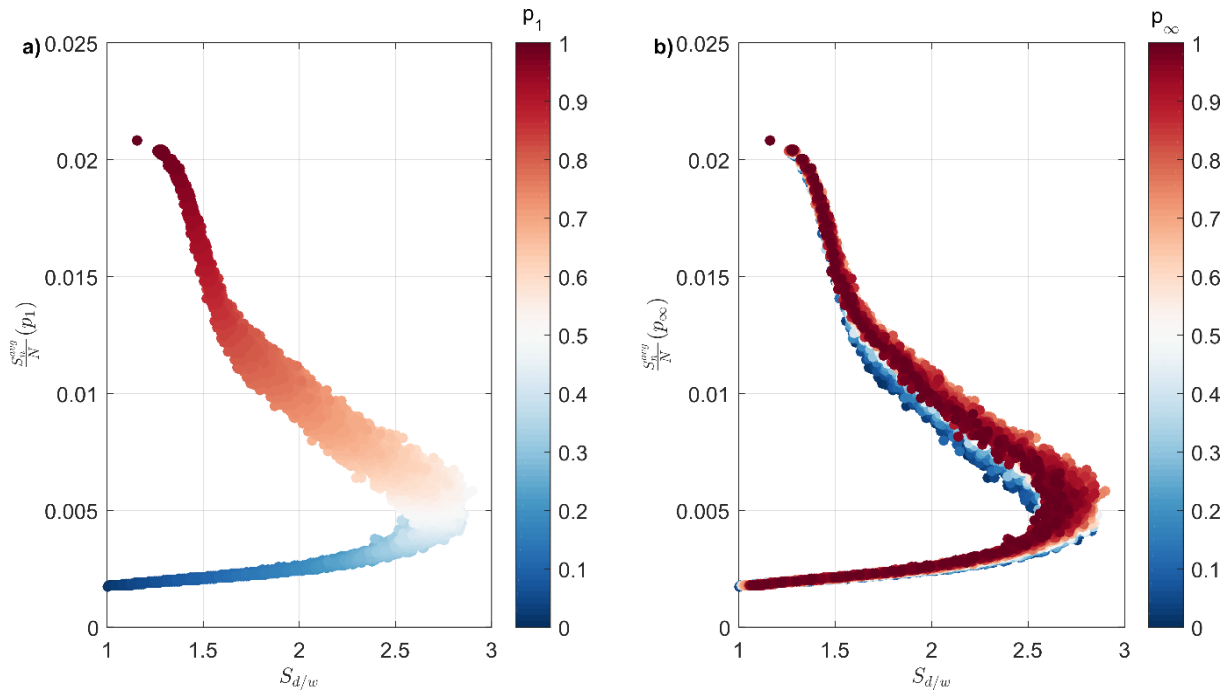

**Supplementary Fig. 4:** Parameter space for the average spreading event size $(S_n^{\text{avg}})$, normalised over the total number of nodes ($N$), and its underlying structure $(S_{d/w})$, as a function of (a) $p_1$ and (b) $p_\infty$.

## 5. Influence of $p_1$ and $p_\infty$ to average propagation rate

Focusing on the relationship between the propagation rate of the largest spreading event, $S_r^{\max}$, and the structure of the pathways used to sustain it, $S_{d/w}$, Supplementary Fig. 5 and Supplementary Fig. 6 map the results obtained under discrete intervals of 0.2 for $p_1$ and $p_\infty$ respectively. Focusing on the effect of indirect exposure, the fork-like behaviour – and the associated presence of two possible trajectories – exist under the limited range of $p_\infty$ values, roughly ranging from 0 to 0.6 (Supplementary Fig. 6). Beyond these values, the two possible trajectories reduce to one, demonstrating the principal role of indirect exposure in controlling the emergence of the two pathways. For comparison, direct exposure has limited control over the emergence of this fork-like behaviour, with $p_1$ being unable to trigger the definitive emergence of the two trajectories (Supplementary Fig. 5).

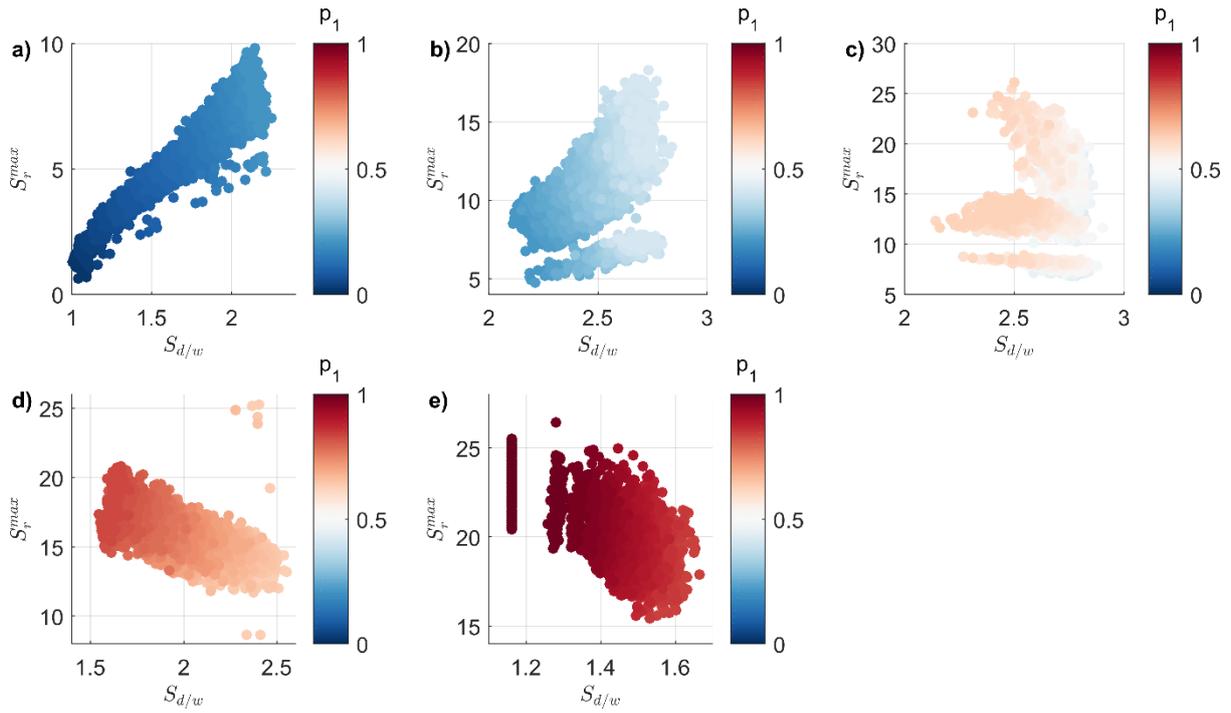

**Supplementary Fig. 5:** Parameter space for the relationship between the propagation rate of the largest spreading event, $S_r^{\max}$, and the structure of the pathway sued to sustain it, $S_{d/w}$, as a function of $p_1$, at 0.2 intervals, ranging from $p_1 \in [0,0.2]$ in subplot (a) – to $p_1 \in [0.8,1]$ in subplot (e).

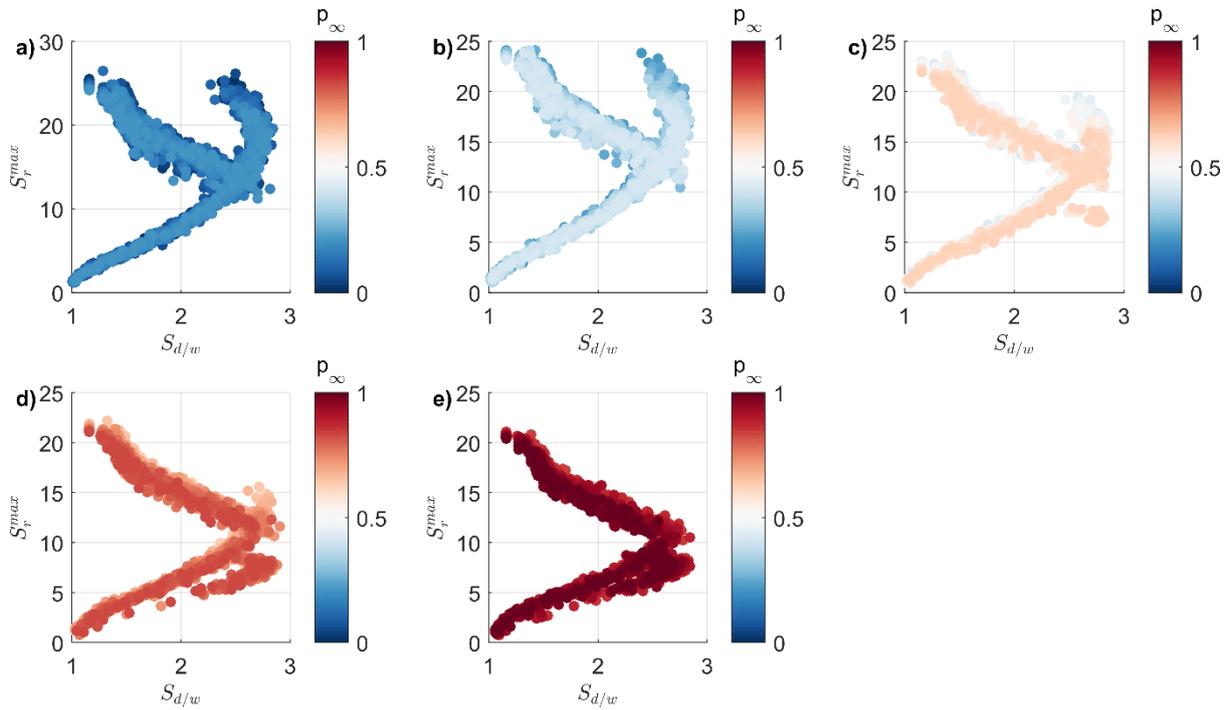

**Supplementary Fig. 6:** Parameter space for the relationship between the propagation rate of the largest spreading event, $S_r^{\max}$, and the structure of the pathway sued to sustain it, $S_{d/w}$, as a function of $p_\infty$, at 0.2 intervals, ranging from $p_\infty \in [0,0.2]$ in subplot (a) – to $p_\infty \in [0.8,1]$ in subplot (e).

## 6. Degree distribution of empirical network

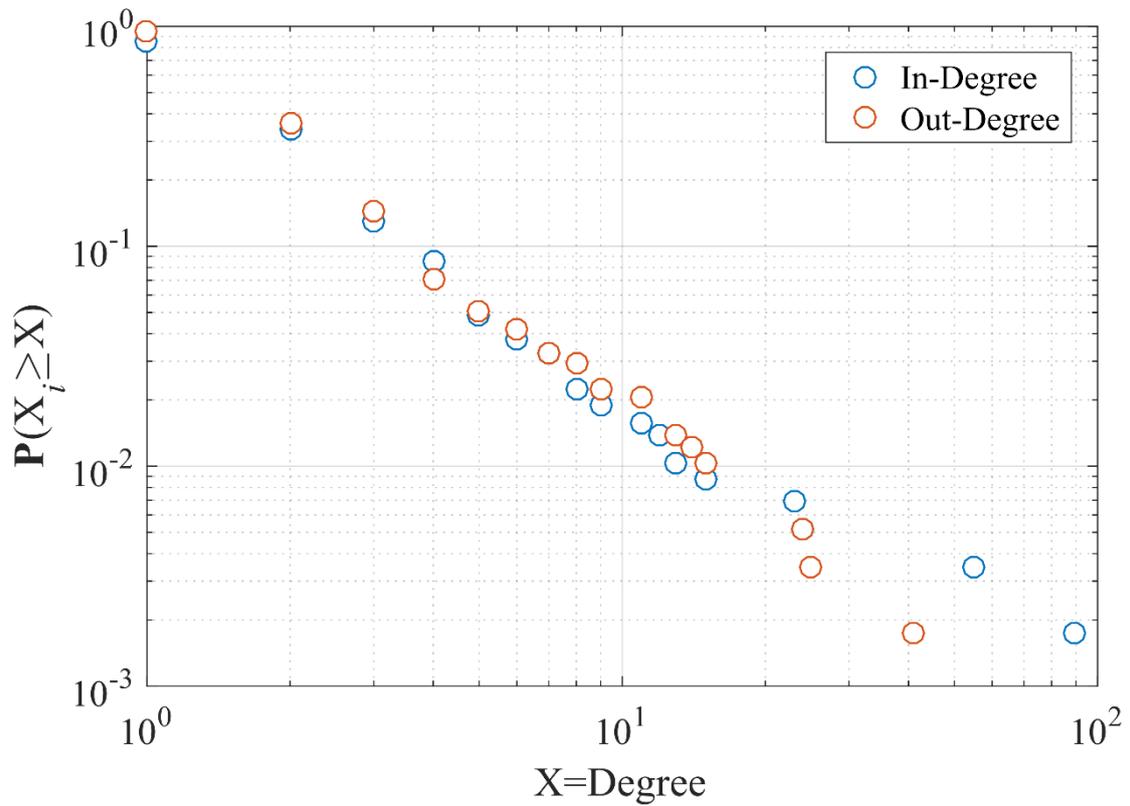

**Supplementary Fig. 7**: Cumulative probability distribution for in-degree (blue) and out-degree (red) of the activity network. Note the heavy-tail nature of both distributions, evident by the straight plot line under log-log axes.